\documentclass[pdflatex,sn-mathphys]{sn-jnl}

\jyear{2021}%

\begin{document}

\title[Soft robotic shell with active thermal display]{Soft robotic shell with active thermal display}


\author*[1,4]{Yukiko Osawa}\email{yukiko.osawa-akiyama@aist.go.jp}
\equalcont{These authors contributed equally to this work.}
\author[2]{Yuho Kinbara}
\author[2]{Masakazu Kageoka}
\author[3]{Kenji Iida}
\author*[1]{Abderrahmane Kheddar}\email{kheddar@lirmm.fr}
\equalcont{These authors contributed equally to this work.}

\affil*[1]{CNRS-University of Montpellier, LIRMM, Montpellier, France}
\affil[2]{Mitsui Chemicals, Inc., Tokyo, Japan}
\affil[3]{Mitsui Chemicals Europe, D\"{u}sseldorf, Germany}
\affil[4]{National Institute of Advanced Industrial Science and Technology (AIST), Japan}



\abstract{Almost all robotic systems in use have hard shells, which is limiting in many ways their full potential of physical interaction with humans or their surrounding environment. Robots with soft-shell covers offer an alternative morphology which is more pleasant in both appearance and for haptic human interaction. A persisting challenge in such soft-shell robotic covers is the simultaneous realization of softness and heat-conducting properties. Such heat-conducting properties are important for enabling temperature-control of robotic covers in the range that is comfortable for human touch. The presented soft-shell robotic cover is composed of a linked two-layer structure: (i) the inner layer, with built-in pipes for water circulation, is soft and acts as a thermal-isolation layer between the cover and the robot structure, whereas (ii) the outer layer, which can be patterned with a given desired texture and color, allows heat transfer from the circulating water of the inner part to the surface. Moreover, we demonstrate the ability to integrate our prototype cover with a humanoid robot equipped with capacitance sensors. This fabrication technique enables robotic cover possibilities, including tunable color, surface texture, and size, that are likely to have applications in a variety of robotic systems.}

\maketitle

\section*{Introduction}
Shells of robots used in industry are designed to cover internal mechanical structures (joints and linkages) as well as actuators, sensors and associated electronics and wiring. 
Such shells are also carefully designed to convey the manufacturing signature, i.e., the fabric mark. The trademarks of most existing robotic arms can be recognized at first sight. In manufacturing, robotic shells also withstand environmental influences, such as unexpected shocks or welding sparks, dust, and high humidity. 
Service robots have different requirements and different morphologies and shapes~\cite{sakagami2002iros,wada2008embm,geva2020nature}, yet have covering shells that serve almost the same purpose as those in industrial robots. 
In the late 19th century, capabilities in terms of tactile sensing, located at the gripper “fingertips'', were investigated to increase robotic manipulation dexterity. With the rise of cobotics (robotics for ``collaborative robots'' called cobots), tactile sensing has been extended to the entire “robot body”, and several solutions have been proposed to make robots aware of external contact or touch; some of these solutions involve direct sensing using sensitized shells (called skins)~\cite{bartolozzi2016nature,tomo2018ral,boutry2018SciRobo,yamane2019humanoids,ma2020nature,bergner2020sensor} and others involve indirect sensing by monitoring a subset of robot state residuals~\cite{haddadin2017tro}. However, there does not appear to be a reason for a robotic shell to be systematically rigid. Robots can instead be covered by soft shells. Yet, why soft shells?

In fact, no large terrestrial moving animals with highly rigid shells (conceived by nature as exoskeletons) exist. 
Large terrestrial animals with rather rigid endoskeletons (bones) surrounded by soft matter (muscles, veins, skins...) have evolved in nature because such a structure gives vertebrates the equivalent of airbag protection in a car. 
Moreover, the soft-shells of relevant living species not only absorb impacts and allow dynamics through self-manipulated motion but can also be used to interact with the external environment; for instance, certain soft-shell animal skins can absorb liquids~\cite{shrestha2018science}, exchange thermal energy~\cite{rushmer1966science}, give a soft impression~\cite{arnold2017soro}, and provide a pleasant feeling to others~\cite{hayward2013science}. In addition, many animals convey information that is useful for social interactions through their skin (e.g., chameleons change their skin color to protect themselves and express their feelings~\cite{teyssier2015science}). 
Such features can be applied to the robotic cover, expanding the possibilities to increase its capabilities in not only sensing, safety and communication functionalities but also tactile and visual aesthetics. 

In particular, thermal impression in physical human-robot interaction is strongly related to human feeling such as sense of trust and friendship toward robots~\cite{nie2012hri,park2014robotica}, secure and comfort feelings~\cite{olausson2014neuroscience,rolls2008neuroimage}. 
We believe this feature is essential for a close human-robot relationship, and we aim at investigating soft shells with thermal capabilities for robots. 
The capability for rendering intended thermal stimuli is called thermal display, developing in telepresence haptics research~\cite{monkman1993tra,drif2005irs,jones2008haptics,guiatni2008springer,jones2016presence,sato2016haptics,osawa2018tie,kajimoto2019haptics} and virtual reality~\cite{ino1997roman,kron2003haptics,kheddar2003mhs}. Most of these studies use a thermoelectric module (called a Peltier device) as a thermal display, which can induce a temperature difference between the front and back sides of the device by means of the so-called Peltier effect~\cite{rowe2018thermoelectrics}. Human-in-the-loop control strategies for the Peltier device have been investigated in many past studies and formulated as temperature-control problems~\cite{yamamoto2004icra,citerin2006icra,jones2008presence,jones2007tap,osawa2016jia}, heat-flux control problems~\cite{guiatni2009presence}, or a mixed temperature/heat flux bilateral-control problem~\cite{drif2005irs}. However, the surface of a commercially available Peltier device is rigid because it is covered by hard ceramics. There are also flexible heaters and thermoelectric modules based on printed electronics techniques~\cite{ding2020nature,yong2018amp}, not yet applied to robotics. For instance, the stretchable resistive heater using Joule effect in~\cite{kim2020jmca} uses air cooling. The skin-like thermo-haptic devices~\cite{lee2020adfm,lee2020adfm2}, targeted for virtual reality applications, have both heating and cooling capabilities and are potentially very promising technologies. For the time being, their use in robotics would require an additional system to radiate the generated heat. Also the wiring might be an issue for large surfaces. In addition, the tactile impression of the surface tends to harden to some extent due to the semiconductor and liquid metal printed (or vapor deposited) on the modules.   

Moreover, few materials exist that can simultaneously achieve the antagonistic characteristics of relatively low stiffness and high heat conductivity. That is, although lower density materials (e.g., silicone rubber and foam) can be used to design soft shells, heat transfer occurs by the motion of electrons and phonons in high-density materials. Thus, because of their low densities, most soft materials have no or very poor capability to be turned into thermal displays~\cite{yamane2018robosoft,ishiguro2007springer}. The few robotic systems embedded with thermal displays~\cite{pena2020hri} (or temperature control systems~\cite{inaba2020ral}) tend to have rigid shells. 

To the best of our knowledge, no soft robotic cover exists that realizes ``display'' features that can both (i) serve aesthetic purposes (e.g., different colors and texture patterns are already possible on rigid covers, and extend to soft covers) and (ii) be tuned with temperature to provide pleasant human touch without compromising other sensing technologies. Assembling or embedding a robotic soft cover with distributed small elements of Peltier devices is not conceivable. Obviously, the surface cannot be uniformly soft, and energy-wise it is certainly not a viable option. Overcoming these shortcomings, our robotic cover can display temperature using a closed-circuit water-circulation system inspired from human blood circulation. The proposed design, materials, and system construction achieve thermal rendering while maintaining a uniformly soft and smooth texture cover. 
The presented soft-shell robotic cover involves a combination of a foam layer and a gel layer with closed-circuit channels for water circulation. A novel water-circulation system is also developed, through which both heating and cooling of the cover are achieved within short response times. The cover was developed considering four kinds of features related to pleasant touch: stiffness~\cite{pasqualotto2020nature}, surface texture patterns~\cite{hayward2013science,hayward2011haptics}, color~\cite{valdez1994psyc}, and temperature~\cite{olausson2014neuroscience,salminen2013springer}. Comparative studies are presented in Table~\ref{tab:aesthetics_comparison}. Moreover, the thickness, stiffness, roughness, and color of the proposed cover can be customized and optimized at will, according to the user or application demands. Finally, we test our proposed cover on the upper right arm of the humanoid robot HRP-4. Capacitance experiments demonstrate the ability to integrate a prototype cover with a humanoid robot equipped with capacitance sensors.  

\section*{Results}
\subsection*{A novel soft cover design with thermal conductivity}
\begin{figure}[t!]
\begin{center}
\includegraphics[width=\columnwidth]{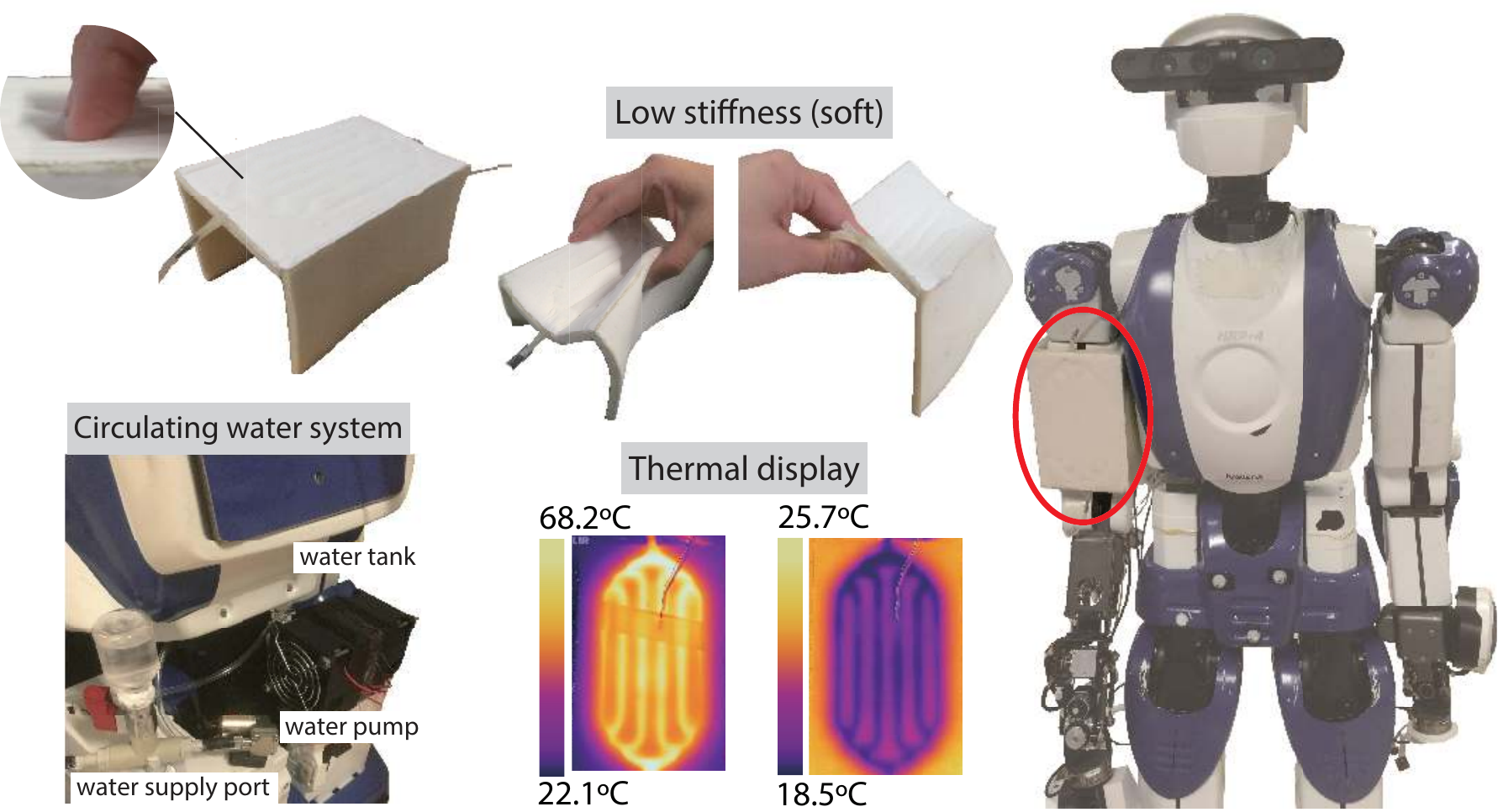}
\caption{{\bf The developed robotic cover. } 
The robotic prototype cover is designed to fit the rigid shell of the humanoid robot HRP-4 right upper arm; it has a $\sqcup$-shaped to cast the upper arm. Thanks to the materials we used, the cover is deformable that can give a soft tactile impression. The water flow channels are built on the upper arm front side (contactable part by a human). The sides parts of the $\sqcup$-shaped cover are left without heating for comparison purpose (see later the capacitive sensing part). The circulating water system for controlling the cover temperature was mounted on the HRP-4 backside.}
\label{fig:robotic_cover}
\end{center}
\end{figure}

Our robotic shell design is inspired in part by the human skin and blood circulation veins~\cite{parsons2014}. Blood flow under the skin surface plays a vital role in maintaining the balance between heat production and heat loss in many living organisms. The blood vessel diameter changes according to the external environment temperature to regulate the amount of blood flow. By analogy, we developed a water-circulation system using built-in pipes in composite soft materials, constituting a robotic cover (see Fig.~\ref{fig:robotic_cover}). Our system includes temperature-controllable heat sources and a mini water-pump to regulate the water flow. The soft cover consists of two main layers. The inner layer is made of low-density foam in which the pipes are patterned as semicircular conduits for water circulation. The outer surface layer, assembled on top of the previous layer, is made of a gel mixed with boron nitride (BN) powder, which is covered by a polyurethane (PU) coating (see the next section for details). The inner foam material has low density, which means low stiffness (compliance with contact) and lightweight (to cover large areas of robots such as humanoids) but very low heat conductivity (isolation). Since the foam layer accounts for a major part of the cover, the thickness of the foam should be reasonable to avoid hindering robot kinematic motion ranges (6~mm for our prototype). The outer cover has higher heat conductivity relative to the inner part; hence, it also has a higher mass density. The thickness of the outer cover should be as thin as possible (we chose 1~mm). The thicknesses of the two layers can be tuned based on the robot design requirements and constraints (see Table~\ref{tab:physical_properties}). The built-in pipes allow heat transfer from the water circulating inside the shell to the robot surface through the outer layer. This solution permits higher performance and lower fabrication costs with respect to embedding separate pipes, as in our preliminary trials~\cite{osawa2020iser}. Because separate pipes add an additional layer to heat transfer and make the surface harder, we challenged not using pipes as a separate component. 

Our prototype is made to fit the upper arm of a humanoid robot we have because (i) it is already embedded with capacitive sensing (this is not the case for the forearms, forelegs, hands, and feet), (ii) the cover prototype can be easily mounted and unmounted for trials. The final objective is covering the humanoid or any other collaborative robot body entirely. The system is thought to provide a complete solution to physical interaction for assistive purposes (see Figure~S1 in supplementary material). The study in~\cite{bolotnikova2020ral} also highlights use-cases of a humanoid-human assistance considering physical multi-contact with the robotic parts such as the upper arm and interactions that can occur on whole-body. 

\subsection*{Material composition of the cover}
The cover inner layer is made of molded PU foam, which is known for having a cushioning property and durability. This foam is used as a cushion in car seats, armrests in chairs or sofas, etc. This inner layer needs to enable the prevention of channel crushing under pressure in human interaction while keeping the surface soft and deformable. We chose 5~C hardness considering the balance between softness for tactile impression and hardness for maintaining the channel shape. 

The cover outer layer is made of a gel mixed with BN powder known to have high heat conductivity and high electric insulation. The gel layer contains 40~pts.mass of BN to 111.67~pts.mass of polymeric components (that is 26.4$\%$ of BN). Thanks to the BN powder, the gel's heat conductivity was improved from 0.2~W/mK (w/o BN) to 0.37~W/mK (w/ BN), being higher than other soft materials while maintaining the cover soft (15~C hardness) (see Table~\ref{tab:heatconductivity}).

The gel layer is made as thin as possible while the surface remains smooth; if the layer is too thin, then the surface becomes rough due to the protrusion of the foam layer channels. In contrast, a thick gel layer could disturb heat transfer from the circulating water. The inner (foam) and outer (gel) layers are stacked using PU resin adhesive so that the attachment part can sustain low stiffness. 

\subsection*{Semi-circular conduits embedded to the foam material}
\begin{figure}[t!]
\begin{center}
\includegraphics[width=\columnwidth]{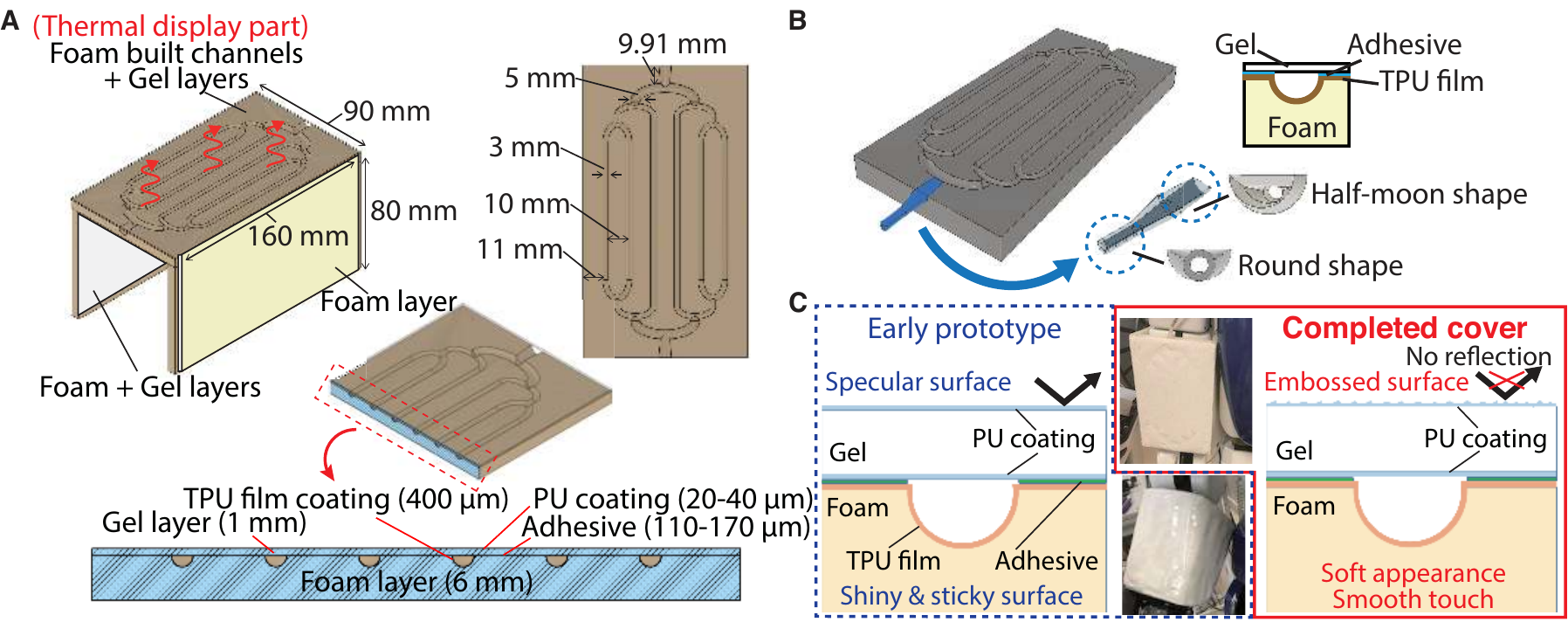}
\caption{{\bf The structure of the robotic cover. } 
({\bf A}) The cover consists of three sides: (1) the foam layer molded channels with the gel layer (proposed one), (2) the foam layer without channels covering the gel layer, and (3) the foam layer itself. The channels embedded with the foam layer are designed as branched into multiple lines to prevent water blockage. 
({\bf B}) Specific connectors are developed by 3D printing for the cover's entrance and exit; the end in the robotic cover side is half-moon shape, smoothly circulating the water through the cover. The semi-circular channels' surface built in the foam is coated by Thermoplastic Polyurethane (TPU) film (400~$\mu$m thickness) with PU coating (20--40~$\mu$m thickness), avoiding permeating the water into the foam layer. 
({\bf C}) Because the gel is transparent, the layer reflects an ambient light; it makes the surface glossy. 
Therefore, the PU coating has an embossed surface to absorb light so that the surface becomes smooth appearance. 
 }
\label{fig:design_structure}
\end{center}
\end{figure}

Since the foam material is liquid before it turns solid, it can be poured into a mold to achieve the desired shape. We designed a pipe pattern on the foam layer, as shown in Fig.~\ref{fig:design_structure}{\bf A}. The semicircular conduits (pipes) were designed to branch into multiple lines, preventing water-flow blockage, e.g., when the robotic arm is firmly grasped by a human. We set the distance (10~mm) and the diameter (3~mm) of each pipe based on the density of the thermal receptors present at the human fingertip (1.6 points/cm$^2$ for warm spots and 2--4 points/cm$^2$ for cold spots~\cite{hensel1973book}).
The pipe surface was coated with a 400~$\mu$m thick thermoplastic polyurethane (TPU) film to avoid water leakage. Moreover, a specific structure of connecting parts was applied to the cover exit and entrance using 3D printing; one end is round, and the other end is half-moon shaped (see Fig.~\ref{fig:design_structure}{\bf B}) to reduce the pressure at the parts flowing into and out of the cover. Both the TPU film and the connecting part can withstand a temperature of 90~$^\circ$C, allowing circulation of up to 90~$^\circ$C hot water inside the cover. 

\subsection*{Surface processing}
In addition to thermal display, the cover's surface can convey other features, such as an engaging appearance, and be patterned with a texture that is pleasant to touch. Because the gel is transparent, the layer reflects ambient light. This makes the surface glossy and unpleasant to both sight and touch. Therefore, the gel layer was coated with PU of 20--40~$\mu$m thickness. As the PU coating has an embossed surface to absorb light, the surface becomes smooth in appearance (see Fig.~\ref{fig:design_structure}{\bf C}). Moreover, the coating protects against corrosion. 

What we propose is a basic design for simultaneously realizing thermal display capability and softness, and the dimensions we chose in this paper are one option among many. To meet varying user or application demands and specifications, the cover thickness, stiffness, surface texture, and color can be customized and optimized at will (see in Table~\ref{tab:customize}). In particular, Mitsui Chemicals has several techniques for post-processing (e.g., coating~\cite{mitsui2018patent}, painting, and cutting), and the tactile impression of the cover surface can be easily changed. 

\subsection*{Robot embedded with a water-circulation system}
We chose water as a circulating fluid because its thermal conductivity is much higher than that of other fluids. For example, water has a thermal conductivity of 0.604~W/mK (21.11~$^\circ$C), whereas other liquids commonly used for cooling have values of 0.05--0.6~W/mK (20~$^\circ$C) (see the detail in Table~S7). Although molten metals have high heat conductivities (e.g., mercury has a value of 8.69~W/mK at 20~$^\circ$C), they are dangerous, inflammable, and require special treatment or conditioning. In contrast, water is ubiquitous, non inflammable, inexpensive, and used in many other heating/cooling systems (including robot actuator cooling~\cite{urata2010iros}). Hence, water is an appropriate liquid for use in our new robotic shell. 

\begin{figure}[t!]
\begin{center}
\includegraphics[width=0.5\columnwidth]{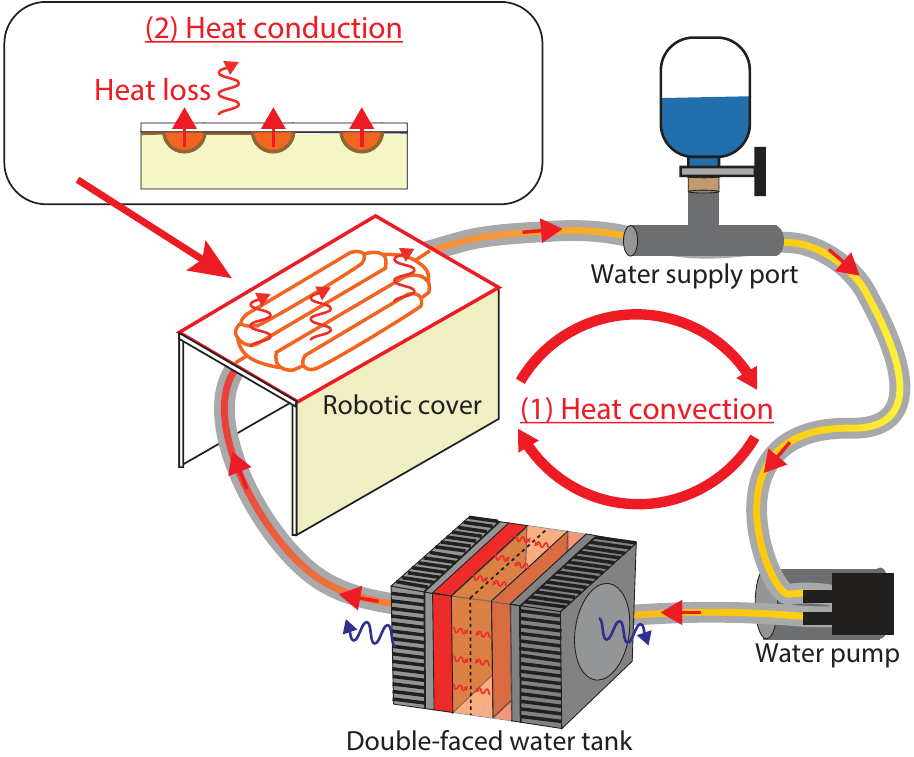}
\caption{{\bf Circulating water system. } 
The circulating water system consists of the robotic cover, water supply port, Peltier devices embedded in the water tank and the micro water pump. 
There are three types of thermal effects in the system: absorbing/generating heat from Peltier devices based on a Peltier effect, heat convection through the circulating water (1), and heat conduction through the surface layer (gel) of the robotic cover (2). 
The water temperature is controlled by Peltier devices so that the robotic cover's temperature matches the desired one.}
\label{fig:circulating_water_system}
\end{center}
\end{figure}

We designed a closed-circuit water-circulation system that can be mounted on any robot (see Supplementary Note~S1, Figs.~\ref{fig:robotic_cover} and~\ref{fig:circulating_water_system} for prototype mounting on an HRP-4 humanoid). One conventional method for achieving the intended temperature is to mix precooled and preheated water using two tanks~\cite{sakaguchi2014springer,goetz2020haptics} (similar to the temperature regulation of tap water and showers). In this case, the system becomes too large to mount on the robot. Therefore, we devised an original small-size water tank (see Fig.~\ref{fig:design_tank}) considering the thermal properties of water, enabling both heating and cooling. The originality of our design lies in the double-faced heating and cooling tank sources; the top and bottom faces are made of Peltier devices themselves so that water can be heated and cooled by direct contact with the Peltier device's ceramics. Additionally, the tank is made of a resin monomer (see the details in Materials and Methods) that inherently has a low heat conductivity (0.15--0.25~W/mK~\cite{garrett1974physics}), preventing heat loss from the tank sides. The installation of a water supply port ensures that air is completely removed from the closed-loop circulating water. The tank, the supply port, and the cover are connected by a silicone pipe (also to avoid heat loss to occur outside the robot shell), the inner diameter and whole length of which are 2.5~mm and 1.46~m, respectively. 

\subsection*{Thermal display capability}
The water-circulation system consists of a micro water pump, a customized water tank, two Peltier devices, and thermocouples (see Fig.~\ref{fig:circulating_water_system}). These components are essential for real-time control of the robot cover temperature. As stated previously, the Peltier devices are mounted on both sides of the water tank, hence each of the other ceramic surfaces is in direct contact with its heat dissipator. Heat is transferred to the cover by means of (1) convection through the circulating water and (2) conduction through the cover surface layer (see Supplementary Note~S2). The Peltier device's temperature and the water pump's output (velocity of the circulating water) are controlled based on a model predictive control (MPC) formulation so that the robot surface temperature tracks the desired one (see Supplementary Note S3). Active temperature control is needed to regulate (in the sense of control) the cover's temperature to any desired one. This process requires decreasing (cooling) or increasing (heating) to reach the temperature of the cover according to the will or preference of the user. In brief, the water needs to be cooled or heated to track the temperature profile. 

\begin{figure}[t!]
\begin{center}
\includegraphics[width=0.5\columnwidth]{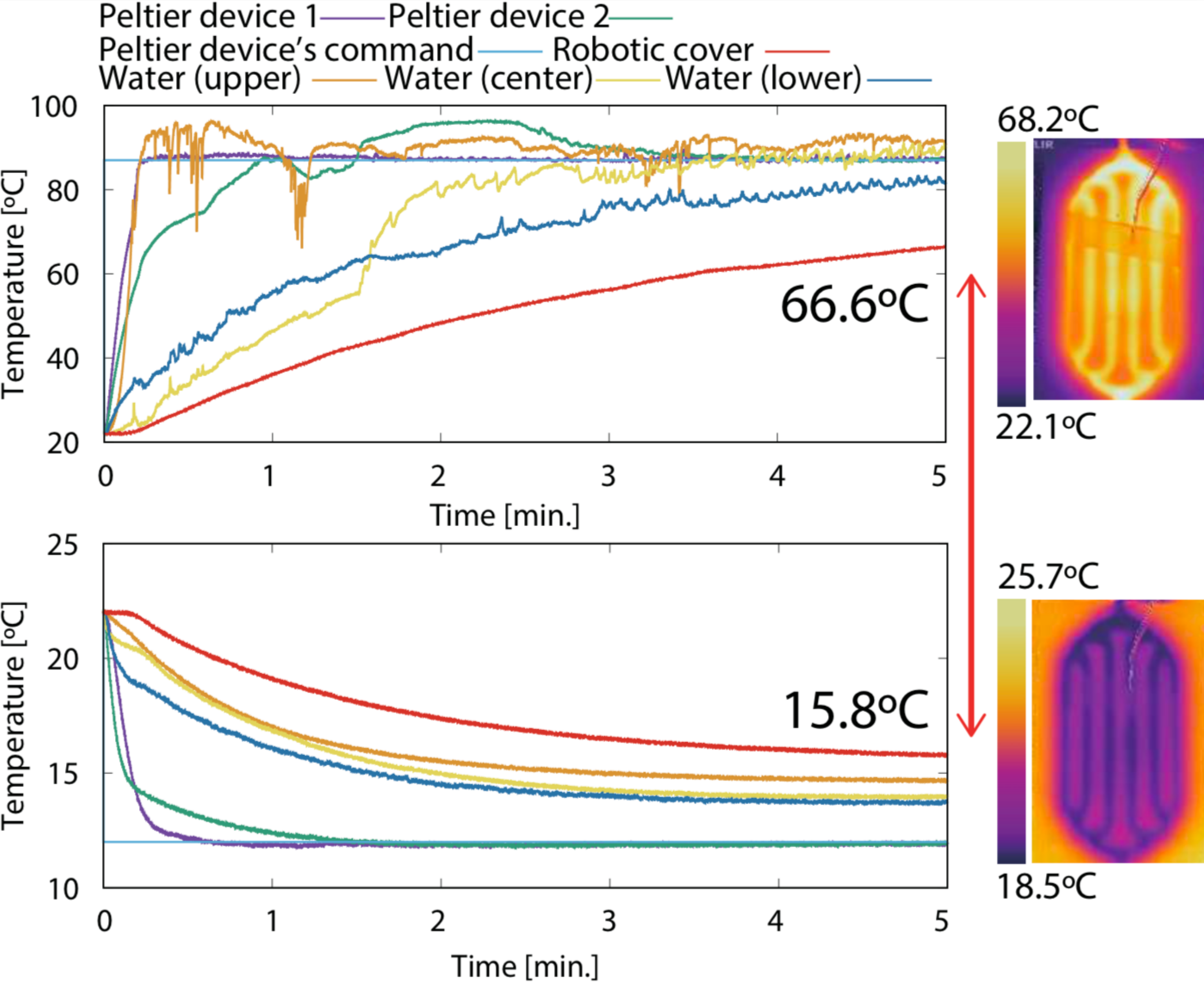}
\caption{{\bf The controllable temperature range of the robotic cover surface. } 
The Peltier devices are controlled to maximum and minimum temperatures in our experimental setup (87~$^\circ$C and 12~$^\circ$C, see Supplementary Note S2). The temperature responses of the water (lower, middle, upper side of the tank) and that of the cover are monitored. 
The robotic cover surface is successfully cooled and heated from 15.8~$^\circ$C to 66.6~$^\circ$C. This range covers the temperature that are comfortable for humans~\cite{olausson2014neuroscience,rolls2008neuroimage,attia1984neuroscience}. 
}
\label{fig:cover_maxmin}
\end{center}
\end{figure}

The cover temperature range can be set according to studies on touch pleasantness found in the neuroscience field. 
According to the literature~\cite{olausson2014neuroscience,rolls2008neuroimage}, humans tend to feel pleasantness induced by hand stimuli at approximately 32--41~$^\circ$C (about 10--15$^\circ$C higher than a standard room temperature). It indicates that humans feel comfortable when they are touched by something with human body temperature. In addition, comfortable temperature changes depend on the environmental temperature: 40~$^\circ$C is comfortable at a room temperature of 20~$^\circ$C and 15~$^\circ$C is comfortable at 40~$^\circ$C~\cite{attia1984neuroscience}. Thus, a pleasant temperature to touch is subjective: it depends on individuals and environmental conditions. Note that the controllable temperature of the cover surface (see Fig.~\ref{fig:cover_maxmin}) includes this range. Considering safety and the electric power system limitation, we set the temperature range of the Peltier device's command between 12~$^\circ$C and 77~$^\circ$C.  

\begin{figure}[t!]
\begin{center}
\includegraphics[width=\columnwidth]{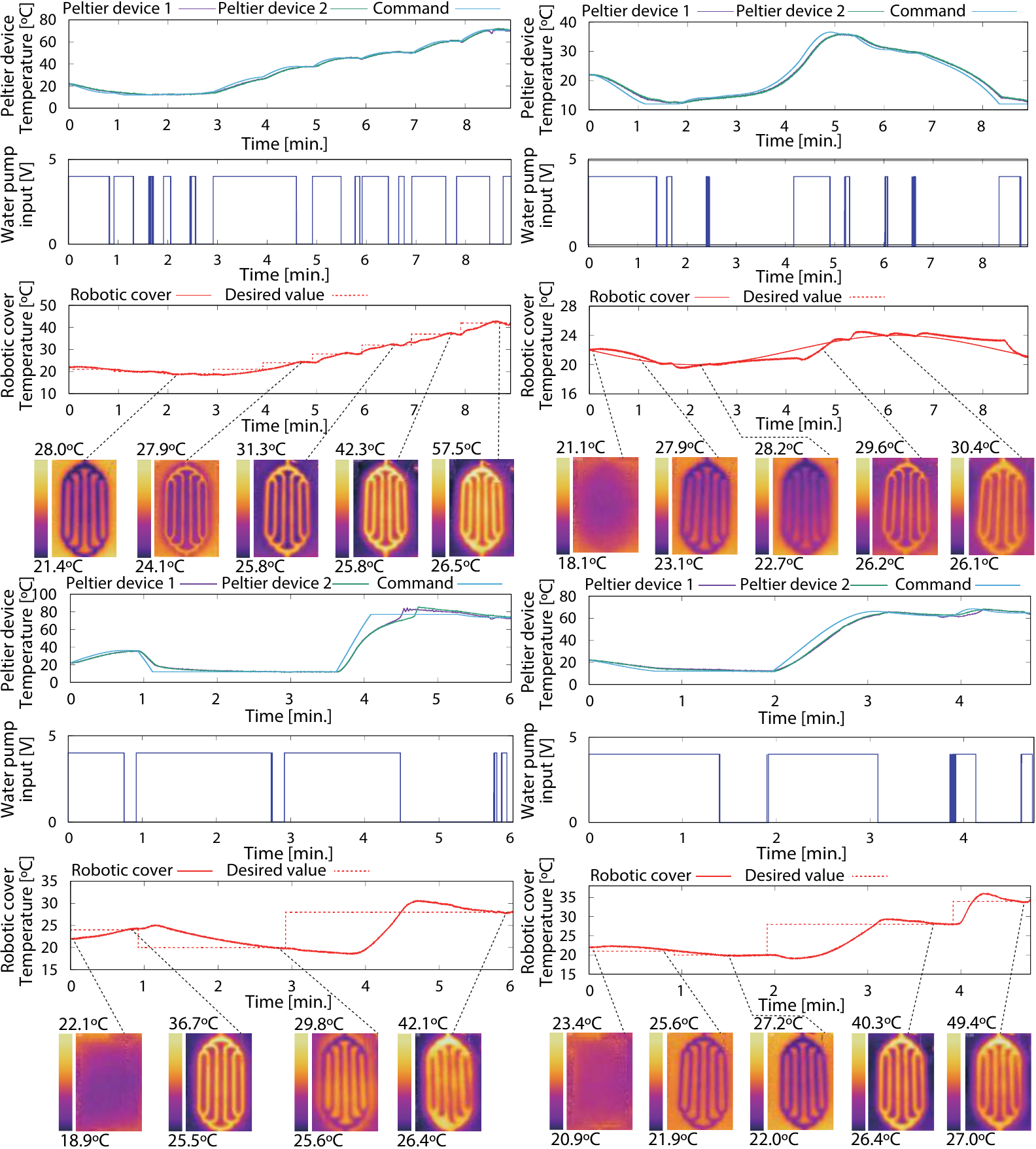}
\end{center}
\caption{\bf The cover temperature control (heat and cool).} The cover temperature controlled with any desired temperature profiles. The cover temperature is regulated to the predefined temperature using the small tank mounted on the humanoid robot. Due to the system's total time constant, the time to reach the command takes more than 1 minute to heat the cooled water and to cool the heated water. This is not critical for human-robot interaction, considering prolonged physical interaction such as hugging and holding. 
\label{fig:tempcon_2}
\end{figure}

We conducted experiments to assess the thermal display capability of the cover using various temperature commands (see Supplementary Note~S4, 5, and 6). Figures~\ref{fig:tempcon_1} and~\ref{fig:tempcon_2} show the temperature responses of the Peltier devices, that of the cover, and the water pump voltage input. The cover temperature was monitored by thermocouples attached to the surface and a thermal imaging camera. The initial temperature was set to room temperature (22~$^\circ$C), measured by an alcohol thermometer.
In Fig.~\ref{fig:tempcon_1}, the cover was heated to 51~$^\circ$C (left figure) and cooled to 17~$^\circ$C (right figure) with step inputs, maintaining each command for 60~sec. The response time for the cover temperature to reach the desired temperature is less than 10~sec. after the water is completely heated or cooled. Mixed temperature profiles with both heating and cooling (rectangular and sine waves) were set as commands in Fig.~\ref{fig:tempcon_2} (Supplementary Movie~S1 shows the results of the lower left figure). These results indicate that the cover temperature can be regulated to the intended value using the closed-loop water-circulation system embedded in the robot. The thermal capability makes it possible to give the desired temperature to users according to environmental conditions and personal preferences. A disturbance from a human hand has little effect on the temperature controller because of the large thermal capacitance of water (see Fig.~S8). 

When cold water was heated or hot water was cooled, the time to reach the command value took longer than that in Fig.~\ref{fig:tempcon_1} for the following reasons. The theoretical values of the water and cover time constants in our system are 32.001~sec. and 5.578~sec., respectively (37.579~sec. in total, see Supplementary Note~S2). Considering the Peltier device time constant and other elements that could disturb heat transfer, the total time constant becomes greater than 45~sec. In addition, the water pump restarted the circulation of the cooled/heated water immediately after changing the command; this additionally cooled/heated the cover, diverting the response away from the command value for a moment. The response is slower than conventional thermal displays using Peltier devices because of the limiting thermal capacitance of water (e.g., the skin-like thermo-haptic device~\cite{lee2020adfm} is 5~sec., whereas our system took from 5 to 60~sec. to reach the temperature commands.) As previously mentioned, using two tanks~\cite{sakaguchi2014springer} is one possible solution to shorten the response time; yet, the system becomes too large to mount on the robot. 

Our cover can offer an additional functionality suitable to the robotics domain: the circulating water system can also be used to transfer and dissipate heat generated by other inside robot features (namely the joint actuators). For example, in humanoid robots, these are made thanks to fans placed near each actuator. These fans generate noise and are not always very efficient. A system cooling in conjunction with our cover can use the same fluid to make it circulated around the actuators for cooling purposes, e.g.~\cite{urata2010iros}. Note that the cover's response speed is not critical for human-robot interaction; the thermal capability is designed for use in prolonged interactions such as holding, considering the frequency range of the human body thermal regulation system, and at providing secure and pleasant expression (For example, a dynamic temperature change (0.5~$^\circ$C/s) is rated as less pleasant than static stimuli in~\cite{salminen2013springer}.) 

We conducted an endurance experiment by operating the system for 7~hours (maximum time allowed in a laboratory with constant monitoring), with a continuous control of the cover temperature to 32~degrees. The water didn't leak from the cover, yet we have witnessed potential amelioration in the connections of other parts of the entire circuit. 

\subsection*{Thermal display with concurrent capacitive sensing}
\begin{figure}[t!]
\begin{center}
\includegraphics[width=\columnwidth]{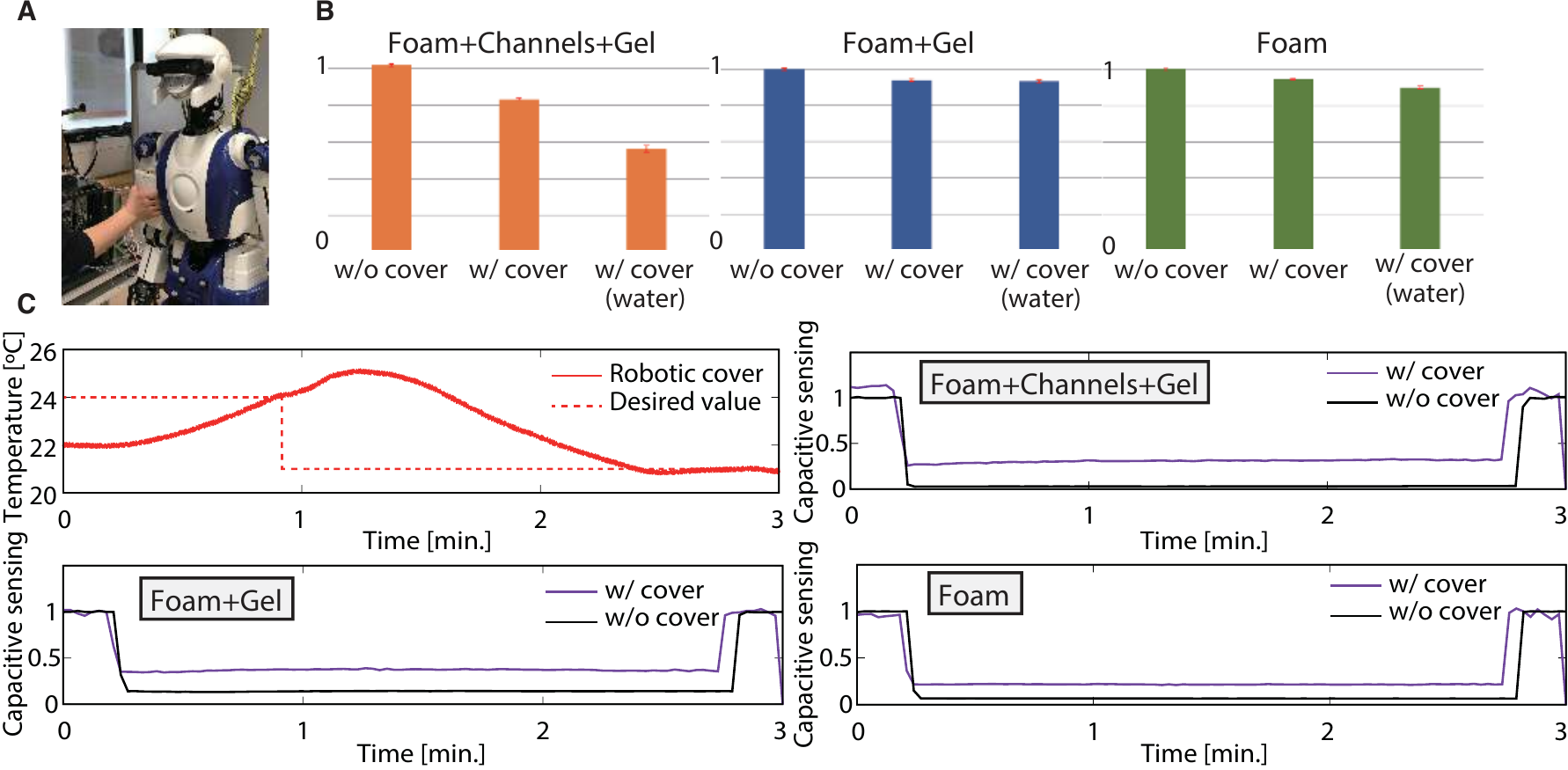}
\end{center}
\caption{\bf Concurrent thermal display and capacitive sensing.}
({\bf A}) We investigated the effect of the cover on humanoid's capacitive sensing by grasping (full touch) the upper arm. 
({\bf B}) The capacitive sensor outputs were compared in three cases: (i) naked arm (without cover), (ii) with the cover empty, and (iii) with the cover filled with water. The three graphs show the average and the standard deviation (red line) of the capacitive sensor's outputs recorded for 20~sec. without contact. 
Namely, the cover affects the capacitive sensing because the cover's material itself and water are conductive. 
({\bf C}) We recalibrated after fitting the cover so that it does not affect the sensing. The outputs were monitored while heating and cooling the cover, grasping in 3~minutes except 20~sec in the beginning and the end. These results suggest that our operating cover can be used concurrently with capacitive sensing. 
\label{fig:capacitive_sensing}
\end{figure}

Proximity sensors~\cite{navarro2021tro}, such as capacitive sensing, can make the robot aware of human closeness prior to contact. This is particularly useful in close human-robot interactions, where robot embedded cameras and existing human tracking fail to provide such information. We embedded our robot with a capacitive sensing system to operate in an industrial setting; see~\cite{kheddar2019ram} for more technical details. We therefore investigated whether our proposed thermal cover can be used concurrently with capacitive sensing. We cast our cover on the right arm of an HRP-4 humanoid embedded with four electrodes for capacitive sensing --one on each side of the arm (front, rear, left, and right sides). Capacitive sensing measures the electric capacitance between the surrounding conductive object and the sensor itself. The raw values are normalized and restricted from 0 to 1; the initial value without contact is set to 1 after calibration and approaches 0 when a conductive material (here, a human hand) approaches the sensor. At this stage, given our optimized prototype cover design, we wanted to evaluate the influence of each layer of our cover on capacitive sensing (this is the very reason it was shaped in $\sqcup$-form). The front part of the cover is the complete cover with the water-circulation pipes (foam+channels+gel, with and without water); it acts on the front capacitive electrode of the arm. The close-to-torso side of the cover is constituted only by the cover outer and inner layers (foam+gel) without the pipe (and hence without water) layers of the cover and act on the left-side capacitive electrode of the arm. The remaining side of the cover is constituted only by foam and acts on the arm right-side electrode. We then investigated and compared the influence of each part of the cover on its associated capacitive electrode (front, left and right sides of the right arm) on each cover side (see Fig.~\ref{fig:design_structure}{\bf A}), fitting the cover on the robot while grasping (full hand touch) it (Fig.~\ref{fig:capacitive_sensing}{\bf A}). Figure~\ref{fig:capacitive_sensing}{\bf B} shows the mean and standard deviation (red line) of the sensor outputs recorded for 20 sec. without contact. Each of the cover side (foam+channels+gel, foam+gel, and foam) outputs in the three cases (naked arm and fitting the cover when it is empty inside and filled with water) are summarized as follows:
\begin{itemize}
\item naked arm (w/o cover): 1.00  (standard deviation:~4.49$\times 10^{-3}$)
\item foam: 0.946 (standard deviation:~3.21$\times 10^{-3}$)
\item foam+gel: 0.940 (standard deviation:~4.88$\times 10^{-3}$)
\item foam+channels+gel: 0.833 (standard deviation:~5.06$\times 10^{-3}$)
\item foam+channels+gel+water: 0.563 (standard deviation:1.67$\times 10^{-2}$)
\end{itemize}

The cover affects statically capacitive sensing because the cover material itself and water are conductive. 
Thus, a recalibration after fitting the cover was performed to get rid of subsequent off-set, as shown in Fig.~\ref{fig:capacitive_sensing}{\bf C}. The outputs were monitored while heating (controlled to 24~$^\circ$C for 1 minute) and cooling (controlled to 21~$^\circ$C for 2 minutes) the cover, with grasping (full touch of the arm) for 3~minutes in total except for 20~sec. in the beginning and at the end. These results suggest that the temperature and water circulation do not affect capacitive sensing, allowing integration with the robot cover (Supplementary Movies S2 and S3 show examples for integrating our shell to the HRP-4). In each of these figure parts, an offset occurs with respect to the naked arm (w/o cover) because the cover thickness does not allow the electrodes to touch, as is the case in the bare (w/o cover) scenario.

\section*{Discussion}
Most recent robotic shells have been developed with functional features such as sensing, impact absorption, and dust-or waterproofing. Additionally, the robot shell can also play a major role in social interactions. We propose a novel design for a lightweight robotic shell (i.e., cover) that is both soft and can be cooled or heated at will. Moreover, our cover can be patterned, colored, and optimized in stiffness and thickness. Our main contributions are the followings:
\begin{enumerate}
\item The soft cover has built-in pipes for water circulation, allowing desired changes in the robot cover temperature. The pipes (or channels) are built-in the inner part of the two-layer structure composing the cover; the material properties (thickness, stiffness...) in each layer can be chosen to fit user or application requirements. 
\item We devised a closed-loop water-circulation system that can be embedded in a humanoid (or other) robot. The developed double-faced tank enables heating and cooling of the water directly and quickly by directly monitoring the water temperature. 
\item We also conceived a control system for thermal rendering considering the robotic cover thermal properties. 
\end{enumerate}
The proposed design has been developed considering four kinds of features related to pleasant touch in human-robot physical interaction; 
 \begin{itemize}
 \item Stiffness: 5~C hardness (inner layer), 15~C hardness (outer layer) (see Table~\ref{tab:physical_properties})
 \item Surface texture patterns: smooth texture and appearance (see Fig.~\ref{fig:design_structure}{\bf A}, {\bf C})
 \item Color: white (the material's original color) for soft and natural impression (see Fig.~\ref{fig:design_structure}{\bf C})
 \item Temperature: actively controllable from 15.8~$^\circ$C to 66.6~$^\circ$C (see Figs.~\ref{fig:cover_maxmin} to~\ref{fig:tempcon_2})
 \end{itemize}
In fact, the cover can be designed by varying several features and parameters. A thermal display on a robotic shell allows using robots for various applications where human trust is essential to promote interaction and acceptance. 

Moreover, our cover does not hinder or compromise capacitive sensing (see Results section), which suggests that it can be applied to existing robots or machines that are equipped with capacitive sensing. Combined with other sensors, it can supply thermal information for superordinate-level techniques (detecting human interactions while providing a comfortable feeling, recognizing humans or materials that contact the cover, etc.). The cover design concept will also be extended to eventually embed other sensing modalities. It can be substantially enhanced with the promising developments in stretchable electronics~\cite{sim2019science,wen2021nature}: one can easily imagine the possibility offered by on-line change of color or on-line change of texture patterns, together with enhanced tactile sensing modalities. We are also working to supplement our cover with a third inner layer that acts to modulate the robot shell stiffness on-line while having high impact absorption capabilities to increase safety. Investigation of the cover's physical impression in terms of usability and human factors is also part of future work. We believe that our system is an innovative solution for promoting human-robot interaction and bringing technology closer to our daily lives and to safely deploy complex robots such as humanoids in domestic and human assistive systems. 

\section*{Methods}
\begin{figure}[t!]
\begin{center}
\includegraphics[width=\columnwidth]{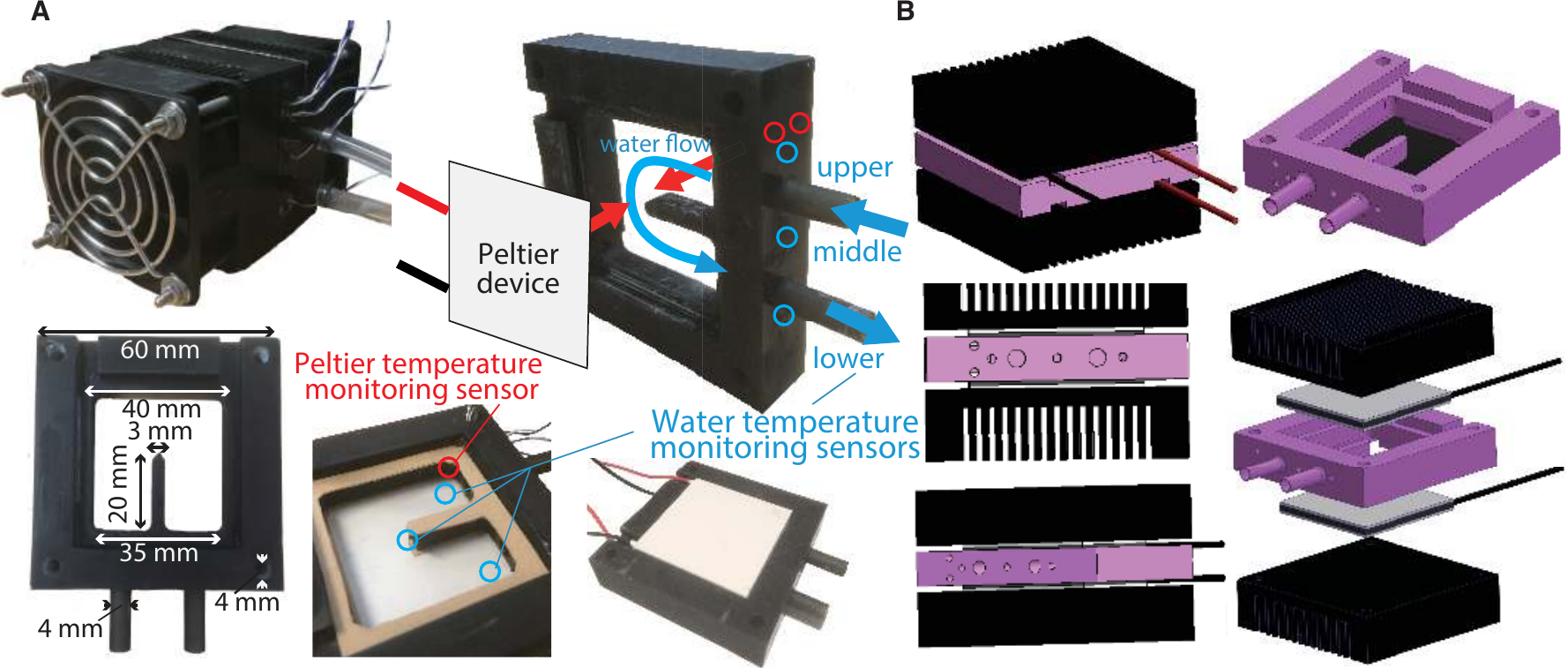}
\caption{{\bf Design of the water tank. }  
({\bf A}) The customized tank comprises two heat sinks integrated with fans, 3D printed tank, and two Peltier devices. The tank has double-faced heat (and cold) sources; the top and bottom faces are made of the Peltier devices themselves so that water can be heated and cooled directly. The gaskets are put between the tank and heat sink to avoid water leakage. There are five holes to attach thermocouples; two are for measuring Peltier devices' temperature, three are for water (upper side, middle, and lower side). The extrusion inside the tank covers the middle sensor for measuring water temperature to avoid contact with the Peltier devices. 
({\bf B}) The water storage part (pink part of this figure) is built by a 3D printer, designed to embed the Peltier devices and sandwich them using heat sinks from above and below. The part is made of a resin monomer, of which heat conductivity is low that can avoid heat loss from the tank wall.  
}
\label{fig:design_tank}
\end{center}
\end{figure}

{\bf The experimental setup}\\
The circulating water system consists of the micro water pump (D200S, maximum flow rate: 80~ml/min., pressure: 11 psi), a 3D printed small water tank (see Fig~\ref{fig:design_tank}), two Peltier devices (ETH-127-14-11-S; maximum temperature difference: 72~$^\circ$C, cooling capacity: 77.1~W, voltage: 15.7~V, current: 8.5~A, area: 40~cm $\times$ 40~cm), and thermocouples ($T$ type, Exposed Junction Wire Thermocouple). The water supplying ports are connected by T-joint connection with 1/8 inch screw port. 
Each part composing the system is connected by a silicone pipe, of which the inner diameter and whole length are 2.5~mm and 1.46~m, respectively. The control system consists of a regular PC with Linux OS and 1.0~msec sampling time, 16~bit AD/DA boards, sensor amplifier (THAB-T-200), and three Micro-Power Motor Amplifiers (TA115, 150~W continuous, 325~W peak). 
\\
{\bf 3D printed part of the water tank}\\
The tank part (the pink part of Fig.~\ref{fig:design_tank}B) was printed by AnyCubic 3D printing (Photon Mono X), 
using Colored UV Resin (color: black, main material: resin and photoinitiator, liquid density: 1.184~g/cm$^3$, solid density: 1.100~g/cm$^3$, hardness(D): 79.0). 
\\
{\bf Thermographic visualization}\\
Thermal images and movies were captured by the thermal imaging camera (FLIR One Pro LT iOS; precision: $\pm$3~$^\circ$C or $\pm$5~$\%$, temperature resolution: 0.1~$^\circ$C.)

\bibliographystyle{sn-basic}
\bibliography{references}

\section*{Acknowledgements (not compulsory)}
The authors would like to thank all colleagues (Tsutomu Tawa, Keiko Sakaguchi, Kazuaki Ichimura, Makoto Kajiura) involved with CNRS-Mitsui Chemicals collaborative work for realizing the robot cover. The authors also thank Olivier Tempier in CNRS-LIRMM France for manufacturing 3D printed water tank. 

\section*{Author contributions}
Y.O. and A.K. proposed the concept design of the robotic cover, Y.K., M.K., K.I., T.T., K.S., K.I., and M.K assisted to realize the design. Y.K. and M.K. developed the cover, K.I. mediated the international collaboration, and Y.O. and A.K. conducted experiments and wrote the initial manuscript. All authors discussed the results and commented on the manuscript.  

\section*{Funding}
This work was supported by the CNRS-Mitsui Chemicals joint research agreement. 
Y.O. acknowledges postdoctoral fellowship support by JSPS Overseas Research Fellowships No.~201960463. 

\section*{Competing interests}
The authors declare no other competing interests. 

\section*{Additional information}
CNRS and MITSUI CHEMICALS, INC. has filed patent applications (Filing No. FR2103637, Filing Date 09/04/2021). 
The corresponding author is responsible for submitting a \href{http://www.nature.com/srep/policies/index.html#competing}{competing interests statement} on behalf of all authors of the paper. This statement must be included in the submitted article file. 
All data needed to evaluate the conclusions are available in the article and the Supplementary Materials.

\begin{figure}[t!]
\begin{center}
\includegraphics[width=\columnwidth]{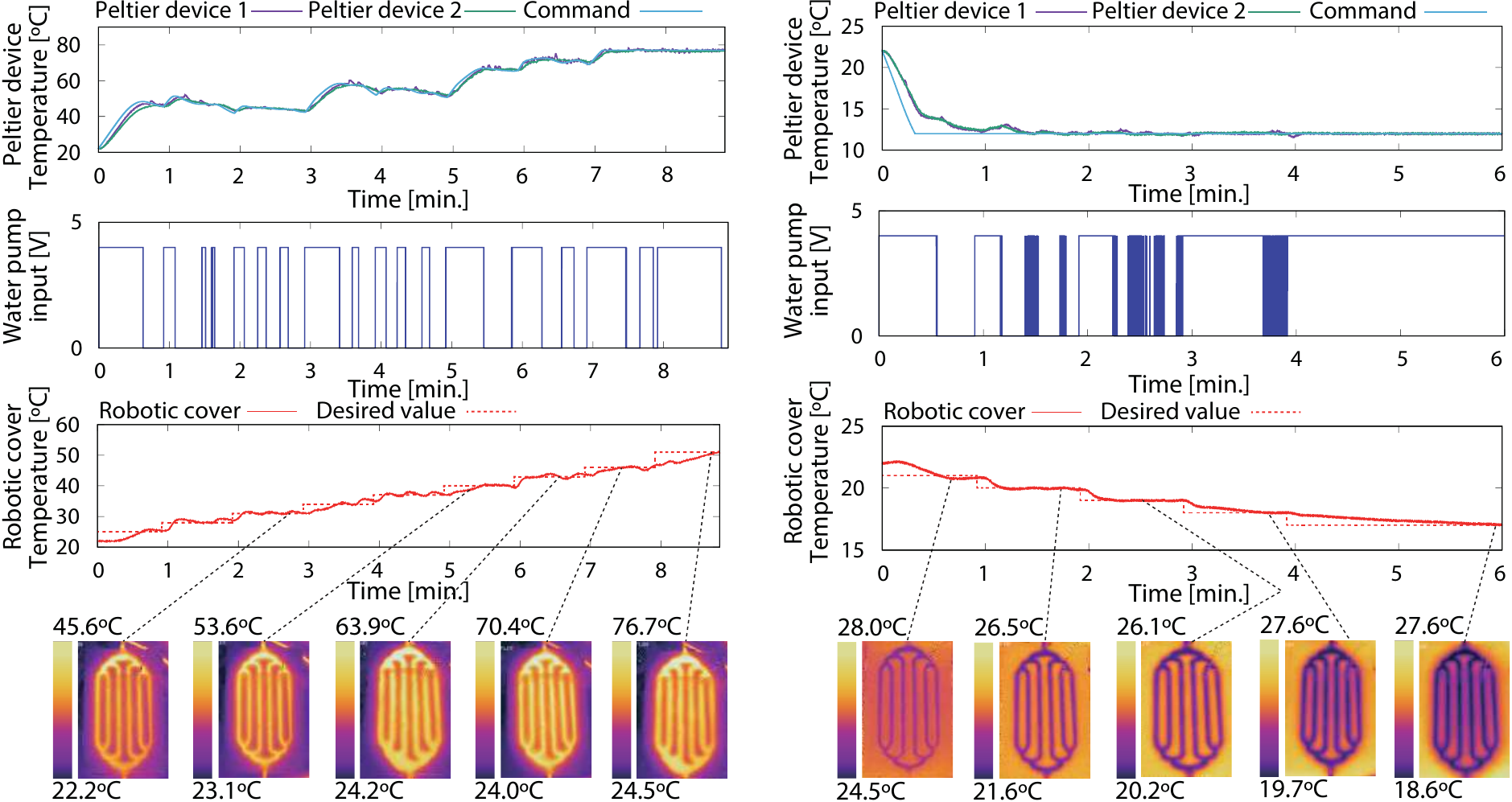}
\end{center}
\caption{\bf The cover temperature control (heat or cool).}
The cover temperature is controlled with successively increasing or decreasing step temperatures. The responses of Peltier device's temperature, water pump's voltage input, and temperature of the cover surface are monitored. The cover temperature is heated by 3~$^\circ$C (5~$^\circ$C in the end) until 51~$^\circ$C (left figure) and cooled by 1~$^\circ$C until 17~$^\circ$C (right figure), keeping each temperature command for 60~sec. Each response speed is less than 10~sec after completely heating/cooling the circulating water. 
\label{fig:tempcon_1}
\end{figure}

\clearpage
	\begin{table}[h!]
		\begin{center}
		\caption{The proposed robotic cover compared with the several studies.} \label{tab:aesthetics_comparison}
		\vspace{5pt}
		\begin{tabular}{c|c|c|c|c}\hline 
			 &stiffness &\fontsize{6pt}{0cm}\selectfont{surface texture} &color &temperature \\
			\hline \hline
			 Soft robot hand and arm system~\cite{yamane2018robosoft} &$\bigcirc$ & & & \\ \hline
			 The android Repliee R1~\cite{ishiguro2007springer} &$\bigcirc$ &$\bigcirc$ &$\bigcirc$ & \\ \hline
			 Musculoskeletal Humanoid Musashi~\cite{inaba2020ral} & & & & $\bigcirc$ (\small heat)  \\ \hline
			 TherMoody~\cite{pena2020hri} & & & & $\bigcirc$ (\small heat $\&$ cool)  \\ \hline
			 The proposed robotic cover & $\bigcirc$ & $\bigcirc$ & $\bigcirc$ & $\bigcirc$ (\small heat $\&$ cool) \\
			\hline 
		\end{tabular}
		\end{center}
		\end{table}

	\begin{table}[h!]
		\begin{center}
		\caption{Physical properties of our robotic cover.} \label{tab:physical_properties}
		\vspace{5pt}
		\begin{tabular}{c|c|c|c|c}\hline 
			\multicolumn{2}{c|}{\small Materials} &\small Thickness & \small Density & \small Hardness~C \\
			\hline \hline
			\multicolumn{2}{c|}{\small Gel layer} & 1~mm & 1200~$\rm kg/m^3$  & 15  \\ \hline
			\multicolumn{2}{c|}{\small Foam layer} & 6~mm & 100~$\rm kg/m^3$  & 5  \\ \hline
			& \small Urethane (surface) & 20--40~$\rm \mu m$ &- &- \\ \cline{2-5}
			\fontsize{6pt}{0cm}\selectfont Coating layer & \small Adhesive & 110--170~$\rm \mu m$ &- &- \\ \cline{2-5}
			& \fontsize{6pt}{0cm}\selectfont{Thermoplastic Polyurethane (TPU) film} & 400~$\rm \mu m$ &- &- \\ \cline{2-5}
			\hline 
		\end{tabular}
		\end{center}
		\end{table}

	\begin{table}[h!]
		\begin{center}
		\caption{Heat conductivity of our robotic cover compared with other soft materials.} \label{tab:heatconductivity}
		The parameters of our cover are measured by the hot disk method (TPA-501, Kyoto Electronics Manuf. Co., LTD, Japan). 
		\vspace{5pt}
		\begin{tabular}{c|c|c|c}\hline 
			 &\fontsize{6pt}{0cm}\selectfont{Heat conductivity} & \fontsize{6pt}{0cm}\selectfont{Measured temperature} & \fontsize{6pt}{0cm}\selectfont{Reference} \\
			\hline \hline
			\fontsize{6pt}{0cm}\selectfont{Our robotic cover (PU gel layer)} & \scriptsize{$0.37$ W/mK} &23 $^\circ$C & Measured*  \\ \hline
			\fontsize{6pt}{0cm}\selectfont{PU gel without boron nitride (BN)} & $0.2$ W/mK &23 $^\circ$C & Measured* \\ \hline
			\fontsize{6pt}{0cm}\selectfont{Silicone rubber} & 0.16 W/mK &25 $^\circ$C & \cite{song2020cej} \\ \hline
			\fontsize{6pt}{0cm}\selectfont{Nitrile-butadiene rubber} & 0.157 W/mK &30 $^\circ$C & \cite{yang2019asm}  \\ \hline
			\fontsize{6pt}{0cm}\selectfont{Natural rubber} & 0.18 W/mK &- & \cite{an2020asm} \\ \hline
			\fontsize{6pt}{0cm}\selectfont{Styrene-butadiene rubber} & 0.232 W/mK &- & \cite{liu2017cst} \\ \hline
			\hline 
		\end{tabular}
		\end{center}
		\end{table}

	\begin{table}[h!]
		\begin{center}
		\caption{The range of the customizable characteristics of the cover provided by Mitsui Chemicals.} \label{tab:customize}
		\vspace{5pt}
		\begin{tabular}{c|c}\hline 
			Characteristics & Customizable range \\
			\hline \hline
			Thickness & Thicker than 1~mm \\ \hline
			Stiffness & Higher than the Asker C hardness 0/0 (Initial/Relaxed) \\ \hline
			Surface texture & Changeable roughness/smoothness by surface coating\\ \hline
			Surface glossiness & Changeable glossy/matte surface\\ \hline
			Color & Changeable by adding pigments or surface coating \\ \hline
			\hline 
		\end{tabular}
		\end{center}
		\end{table}

\end{document}